\documentclass[useAMS,usenatbib]{mn2e}

\usepackage[pdftex,pdfpagemode={UseOutlines},bookmarks,bookmarksopen,colorlinks,linkcolor={blue},citecolor={blue},urlcolor={red}]{hyperref}
\usepackage{graphicx}
\usepackage{caption}
\usepackage{subcaption}
\usepackage{natbib}
\usepackage{rotating,longtable,lscape}

\title[Colour-Magnitude Diagrams of Transiting Exoplanets I]{Colour-Magnitude Diagrams of Transiting Exoplanets\\ I - Systems with parallaxes}
\author[Amaury H.~M.~J. Triaud]
{Amaury H.~M.~J. Triaud$^{1,2}$\thanks{E-mail: triaud@mit.edu}\\
$^{1}$Kavli Institute for Astrophysics \& Space Research, Massachusetts Institute of Technology, Cambridge, MA 02139, USA\\
$^{2}$Fellow of the Swiss national science foundation}

\begin{document}

\date{Accepted ?. Received ?; in original form ?}

\pagerange{\pageref{firstpage}--\pageref{lastpage}} \pubyear{2014}

\maketitle

\label{firstpage}

\begin{abstract}
Broadband flux measurements centred around [3.6 $\mu$m] and [4.5 $\mu$m] obtained with {\it Spitzer} during the occultation of seven extrasolar planets by their host stars have been combined with parallax measurements to compute the absolute magnitudes of these planets. Those measurements are arranged in two colour-magnitude diagrams. Because most of the targets have sizes and temperatures similar to brown dwarfs, they can be compared to one another. In principle, this should permit inferences about exo-atmospheres based on knowledge acquired by decades of observations of field brown dwarfs and ultra-cool stars' atmospheres. Such diagrams can assemble all measurements gathered so far and will provide help in the preparation of new observational programs. In most cases, planets and brown dwarfs follow similar sequences. HD\,2094589b and GJ\,436b are found to be outliers, so is the nightside of HD\,189733b. The photometric variability associated with the orbital phase of HD\,189733b is particularly revealing. The planet exhibits what appears like a spectral type and chemical transition between its day and night sides: HD\,189733b straddles the L-T spectral class transition, which would imply different cloud coverage on each hemisphere. Methane absorption could be absent at its hot spot but present over the rest of the planet.
\end{abstract}

\begin{keywords}
planetary systems -- planets and satellites: atmospheres -- binaries: eclipsing -- stars: distances -- brown dwarfs -- Hertzsprung--Russell and colour--magnitude diagrams.
\end{keywords}


Colour-magnitude diagrams are frequently used in Astronomy to display photometric measurements obtained using different filters. They help identify key features between groups of stars, be it to distinguish different evolutionary stages (e.g. \citealt{Maeder:1974fk}), or select their spectral types (e.g. \citealt{Lepine:2011qy}). In addition to direct imaging \citep{Chauvin:2004lr,Marois:2008fk}, it is possible to measure light re-emitted by an extrasolar planet \citep{Harrington:2006lr}, especially if its orbital inclination produces an occultation \citep{Deming:2005ij,Seager:2010kx}. A subset of occulting systems have had their parallax measured by ESA's {\it Hipparcos} satellite \citep{van-Leeuwen:2007qy}. It is therefore possible to obtain accurate absolute magnitudes for transiting extrasolar planets. Combining measurements in several bands, a colour-magnitude diagram can be assembled. 

Most planetary flux measurements have been acquired using NASA's {\it Spitzer} Space Telescope, notably using the IRAC 1 \& 2 channels, centred at 3.6 and 4.5$\mu$m  \citep{Fazio:2004fk}. These are bands that have frequently been used to observe field brown dwarfs (e.g. \citealt{Patten:2006uq}) and then combined with parallax measurements, to compile colour-magnitude diagrams (e.g. \citealt{Dupuy:2012lr}). Apart from a few, the planets whose flux have been detected fall in the regime of the {\it hot Jupiters}, whose sizes and temperatures are similar to brown dwarfs'. Both population can therefore be compared to each other.

The aim of this letter is to introduce the first colour-magnitude diagrams for transiting planets using the most accurate distances available so far. This is also an opportunity to showcase the interest of producing such diagrams for exoplanetology. 

\section{Procedure}

\begin{figure*}  
\begin{center}  
	\begin{subfigure}[b]{0.45\textwidth}
		\caption{}
		\includegraphics[width=\textwidth]{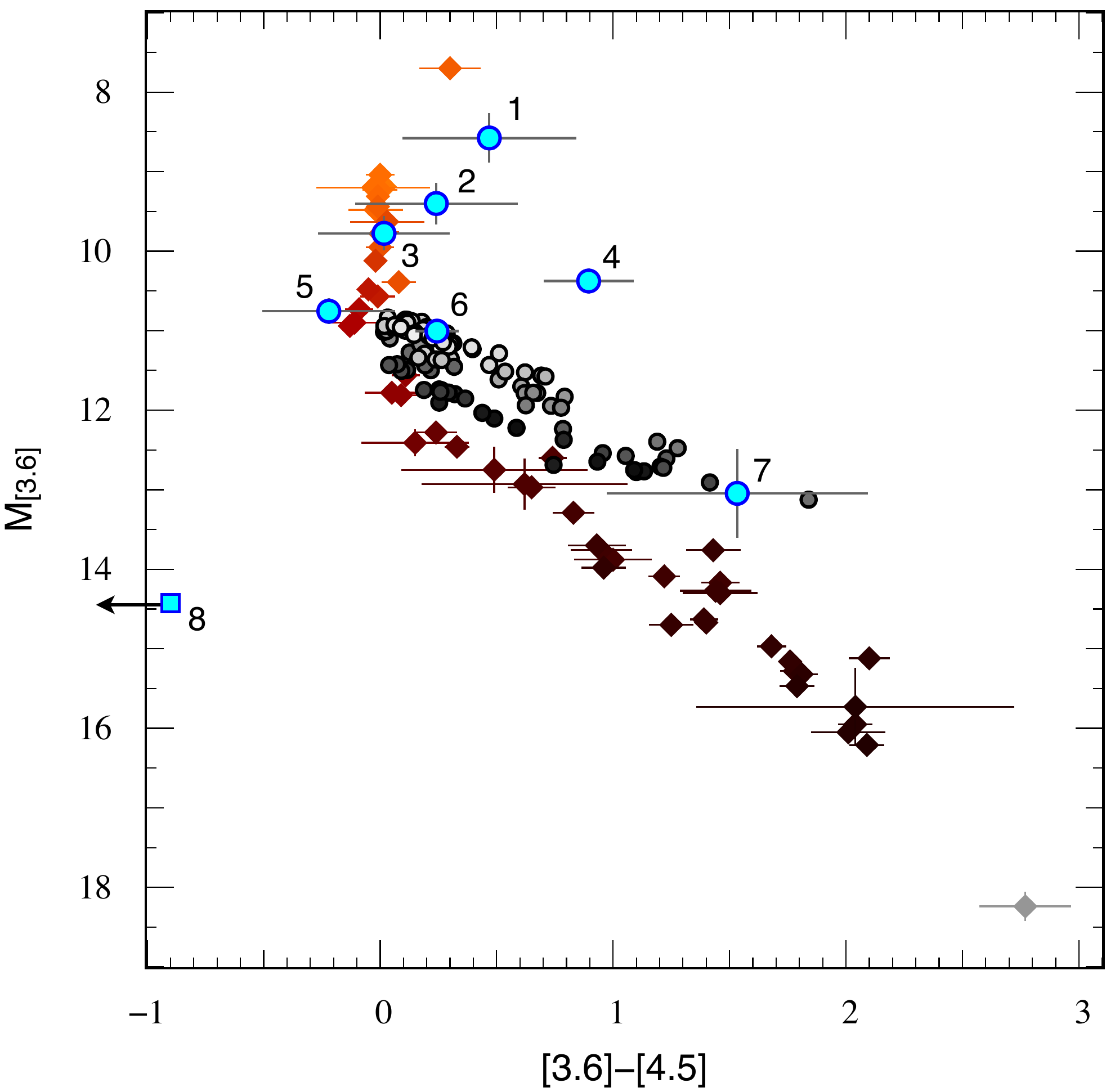}  
		\label{fig:plot3p6}  
	\end{subfigure}
	\begin{subfigure}[b]{0.45\textwidth}
		\caption{}
		\includegraphics[width=\textwidth]{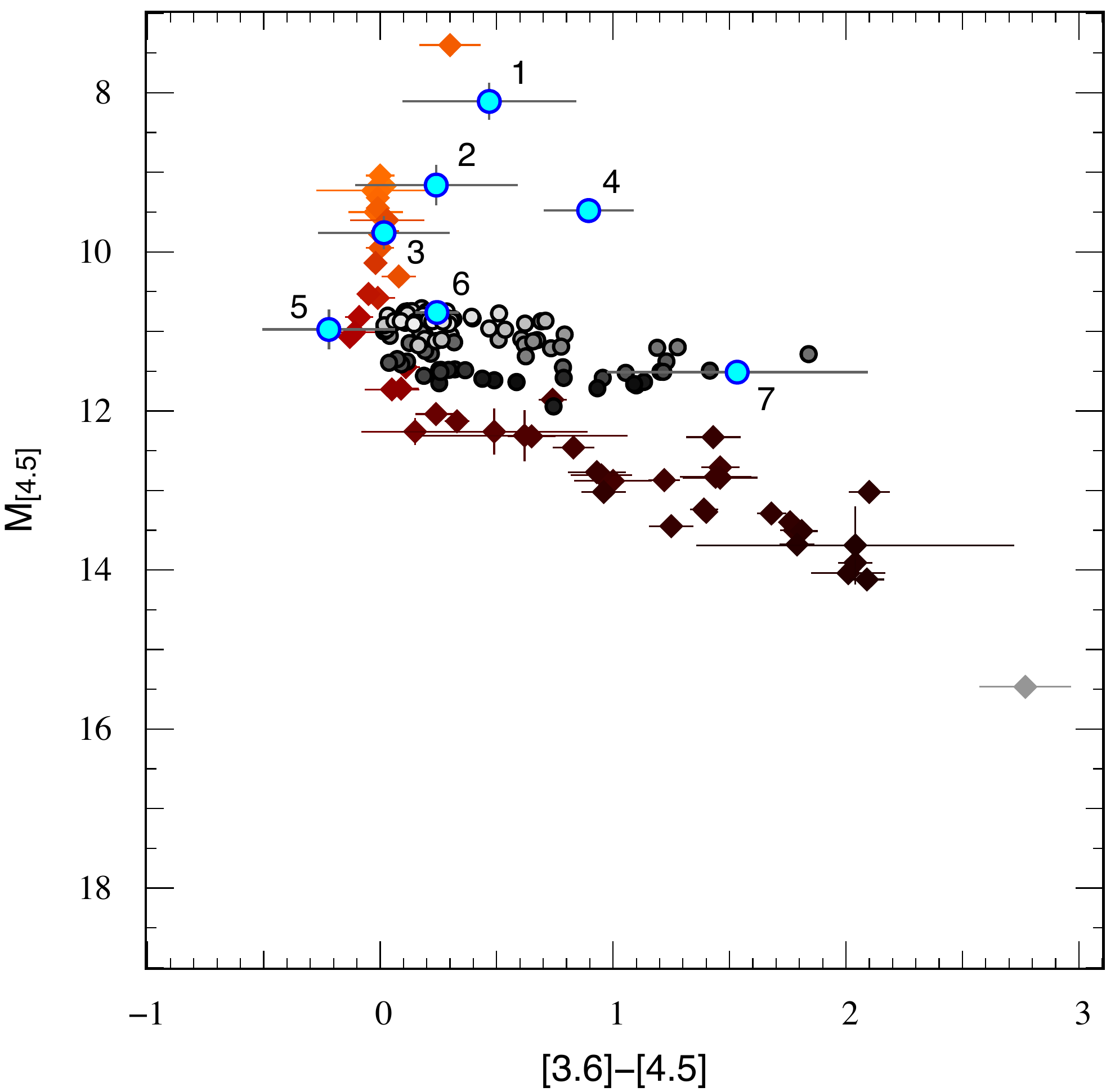}  
		\label{fig:plot4p5}  
	\end{subfigure}
\caption{Two colour-magnitude diagrams of absolute magnitudes at [3.6 $\mu$m ] and [4.6 $\mu$m] versus [3.6]-[4.5] colour. The diamonds show late M dwarfs, L, T, and one Y dwarf as compiled in \citet{Dupuy:2012lr}. The grey diamond is WD\,0806-661B has no spectral type yet \citep{Luhman:2012lr}. The kink at M$_{\rm [3.6]} \sim$11.5 corresponds approximately to a L9 spectral type and the location where methane absorption starts appearing in the [3.6 $\mu$m ] band. Large, blue circles represent the transiting planets whose measurements (and numbers) are located in Table~\ref{tab:planetmags}. Smaller grey circles represent the phase curve of HD\,189733b \citep{Knutson:2012ys}, (lighter gray is for dayside, while darker tones are for nightside). The blue square shows the upper limit of GJ\,436b, which is located to the left of the diagram. 
}\label{fig:plot}  
\end{center}  
\end{figure*}

Searching the literature (e.g. \href{http://exoplanets.org}{exoplanets.org}; \citealt{Wright:2011fj}), there are seven systems observed in both [3.6 $\mu$m] and [4.5 $\mu$m] IRAC channels that also have measured parallaxes in the latest reduction of the {\it Hipparcos} astrometric data  \citep{van-Leeuwen:2007qy}. The systems are listed in Table~\ref{tab:starmags}.  

All those systems have also been recently observed by NASA's {\it WISE} satellite whose W1 and W2 bands are close the IRAC 1 \& 2 channels and whose all-sky photometric catalog is readily available \citep{Cutri:2012fk}. Using parallaxes, absolute magnitudes were computed and errors were propagated (errors are dominated by the uncertainty on the distance, and on the planet's flux). Results are located in Table~\ref{tab:starmags}.


All planetary fluxes were obtained from the published literature and transformed into absolute magnitudes using the standard convention and the values gathered in Table~\ref{tab:starmags}. Most measurements are for the dayside emission. The nightside emission in both channels has only been undoubtedly detected for HD\,189733b. The planet's absolute magnitudes are displayed in Table~\ref{tab:planetmags}. 
They are plotted in Figure~\ref{fig:plot}, overlaid on field brown dwarfs measurements obtained from \citet{Dupuy:2012lr}. Spectral types range from M6 to Y0, encompassing the L \& T classes \citep{Kirkpatrick:2005th,Kirkpatrick:2012lr}

In addition to the flux measurements at superior and inferior conjunctions, \citet{Knutson:2012ys} observed HD\,189733b in both IRAC channels during a full orbital phase. Those fluxes have also been converted to absolute magnitudes and binned for visual convenience. They are represented as grey circles. White corresponds to mid-occultation time (midday), and black to mid-transit time (midnight) as seen from Earth. They are to be read clockwise. 


\begin{table*}
 \centering
 \begin{minipage}{140mm}
  \caption{Apparent WISE 1 \& 2 magnitudes and computed absolute magnitudes for the seven occulting planet hosts having parallaxes.}\label{tab:starmags}
  \begin{tabular}{lrrrrr}
  \hline
   Name	&distance				&         \multicolumn{2}{c}{ Apparent magnitudes} 		& \multicolumn{2}{c}{Absolute magnitudes}\\
   		& (pc)       				& W1 			& W2    			& M$_{\rm[3.6]}$ 	& M$_{\rm[4.5]}$ \\
 \hline
  GJ\,436		&$10.14\pm0.24$	&$5.987\pm0.052$	&$5.703\pm0.025$	&$5.956\pm0.071$	&$5.672\pm0.057$	\\
  HD\,189733	&$19.45\pm0.26$	&$5.366\pm0.065$	&$5.337\pm0.025$	&$3.921\pm0.071$	&$3.892\pm0.038$	\\
  HD\,209458	&$49.6\pm2.0$		&$6.287\pm0.048$	&$6.280\pm0.022$	&$2.808\pm0.098$	&$2.801\pm0.090$	\\
  HD\,149026	&$79.4\pm4.4$		&$6.760\pm0.037$	&$6.804\pm0.021$	&$2.26\pm0.13$	&$2.30\pm0.12$	\\
  WASP-18	&$99\pm11$		&$8.072\pm0.023$	&$8.116\pm0.021$	&$3.09\pm0.23$	&$3.14\pm0.24$	\\
  HAT-P-2		&$114.3\pm9.8$	&$7.562\pm0.025$	&$7.583\pm0.020$	&$2.27\pm0.18	$	&$2.29\pm0.19$	\\
  WASP-33	&$116\pm11$		&$7.429\pm0.027$	&$7.455\pm0.019$	&$2.11\pm0.20$	&$2.14\pm0.20$	\\

\hline
\end{tabular}
\end{minipage}
\end{table*}

\begin{table}
 \centering
 \begin{minipage}{80mm}
  \caption{Absolute magnitudes for the seven extrasolar planet whose hosts have parallaxes. Numbers in the last column correspond to those displayed on Figure~\ref{fig:plot}.}\label{tab:planetmags}
  \begin{tabular}{llrrr}
  \hline
   Name	&				& \multicolumn{2}{c}{Absolute magnitudes} 	& Refs\\
   		&        			& M$_{\rm[3.6]}$ 	& M$_{\rm[4.5]}$ 		& \\
 \hline
  WASP-33b	&day&$8.58\pm0.30$	&$8.11\pm0.22$	& 1\\
  WASP-18b	&day&$9.40\pm0.25$	&$9.16\pm0.24$	& 2\\
  HAT-P-2b	&day&$9.78\pm0.20$	&$9.76\pm0.19$	& 3\\
  HD\,209458b	&day&$10.38\pm0.15$	&$9.48\pm0.12$	& 4\\
  HD\,149026b	&day&$10.76\pm0.15$	&$10.98\pm0.24$	& 5\\
  HD\,189733b	&day&$11.006\pm0.076$	&$10.762\pm0.045$	& 6\\
  HD\,189733b	&night&$13.06\pm0.54$	&$11.51\pm0.12$	& 7\\
  GJ\,436b		&day&$14.42\pm0.11$	&$>15.5$			& 8\\

\hline
\end{tabular}
\\{\bf References:} 1) \citet{Deming:2012rr}; 2) \citet{Knutson:2009fk}; 3) \citet{Lewis:2013qy}; 4) \citet{Knutson:2008qy};  5) \citet{Maxted:2012lr}; 6-7) \citet{Knutson:2012ys}; 8) \citet{Stevenson:2010oq}
\end{minipage}
\end{table}

\section{Discussion}

The most striking feature in the [3.6]-[4.5] colour is  a {\it kink} occurring around M$_{\rm [3.6]} \sim$ M$_{\rm [4.5]} \sim 11.5$ which has been interpreted as the emergence of methane absorption in the [3.6 $\mu$m] band with decreasing effective temperatures \citep{Patten:2006uq}. This kink occurs as brown dwarfs transition from spectral class L to T \citep{Dupuy:2012lr}.

The planets' dayside magnitudes are broadly compatible with the vertical M and L sequence. Two outliers are found: HD\,209458b and GJ\,436b. Most hot Jupiters are bloated \citep{Charbonneau:2000dp,Demory:2011lr}, and thus larger than typical brown dwarfs \citep{Diaz:2013fr,Baraffe:1998ly,Burrows:2011fy}, even under irradiation \citep{Deleuil:2008lr,Anderson:2011fk,Triaud:2013lr}. A change in radius can lead to a change in luminosity. This would make HD\,209458b, $\sim$ 0.9 magnitude brighter and GJ\,436b almost 2 magnitudes fainter than a brown dwarf that has the same effective temperature. This simple correction does not explain alone the discrepancy and suggests the need for  an additional explanation. HD\,209458b's redder colour may point to different atmospheric behaviour than for brown dwarfs and the majority of hot Jupiters. In the current sample this is the planet with the lowest gravity \citep{Southworth:2007qy}. GJ\,436b is also hard to reconcile and may localise a Neptune-sequence, with different chemistry, of which this is the first example. However it is also possible that both measurements have been affected by systematics. For example, GJ\,436b's [3.6~$\mu$m] dayside flux measurement may have been polluted by stellar variability \citep{Stevenson:2010oq}.\\ 

The phase curve of HD 189733b \citep{Knutson:2012ys} is to be read clockwise.  A global offset of the phase curve by -0.5 mag was anticipated due to a radius difference between HD\,189733b and typical brown dwarfs. Starting from occultation (6), the planet cools down and follows a slightly shallower slope compared to the T-dwarf sequence. The reason for this is unclear but could lie in the manner heat redistribution is achieved. As the night-side rolls into view the planet progressively returns to the T-dwarf sequence. It follows the kink to finally rise vertically along the L-dwarf sequence reaching climax at M$_{\rm[3.6]} = 11$ mag. 

Assuming no chemical alteration, an increase in temperature will cause a vertical increase. At [3.6]-[4.5] = 0.5, there is an offset of about one magnitude between the day and night sides. This corresponds to a temperature ratio of ~1.6 between both hemispheres, which is consistent with what has been measured \citep{Knutson:2012ys}. 

The dayside emission, at occultation, is slighter fainter and redder than the brightest emission. The fainting has been already interpreted as a hot spot on the planetÕs atmosphere that is offset from the sub-stellar point by westerly, equatorial winds \citep{Showman:2002qy,Knutson:2007lr}. The vertical rise of the phase curve therefore corresponds to the coming into view of that hot spot. Since it follows a sequence where methane is absent, by comparison, we can infer that there could be no methane absorption at the hot spot, while remaining present in the surrounding regions.

Comparing HD\,189733b's magnitudes to a brown dwarf magnitude-spectral type relation \citep{Dupuy:2012lr}, a spectral type of L5$\pm$2 is expected on the dayside, and a spectral type of T5$\pm$3 on the nightside. The study of the L-T transition is an important research topic. At the transition, clouds that homogeneously cover L-dwarfs start breaking-up, revealing the deeper, hotter part of the atmosphere. This leads to a rapid change in J-H and J-K colours \citep{Tinney:2003uq,Vrba:2004vn,Kirkpatrick:2005th,Saumon:2008fk,Dupuy:2012lr}. HD\,189733b straddles the L-T transition. Future observations of the dayside in J, H and K bands would be particularly interesting to obtain in order to check whether the planet follows a sequence defined by young,  directly imaged planets \citep{Barman:2011vn} or whether it behaves as field brown dwarfs do. If it does, we would expect a cloud cover on the dayside and a partly cleared weather on the nightside. A recent analysis of Kepler-7b's phase curve demonstrated inhomogeneous cloud coverage can exist \citep{Demory:2013uq}. This would imply that the transmission spectrum at ingress is likely different than at egress with repercussions on the current interpretation of such data. 
WASP-33b \citep{Deming:2012rr}  and WASP-18b \citep{Maxted:2012lr} are marginally redder than the empirical M and L sequence. One could conclude their hot spot are also east of the substellar point, which is verified in WASP-18b's case \citep{Maxted:2012lr}. 

Studying Figure~\ref{fig:plot} empirically informs that most {\it hot Jupiters} should be devoid of methane absorption on their illuminated hemisphere since most objects  are located above the kink and follow the vertical M \& L dwarf sequence. More measurements would be needed to observe whether there exists a distinction between the pM and pL classes were proposed by \citet{Fortney:2008lr}. Targets whose dayside emission is slightly cooler than HD\,189733b, such as WASP-80b \citep{Triaud:2013sk}, will be of particular interest as they would presumably fall within the T range.

 If carefully estimated, photometric distances can be accurately determined (e.g. \citealt{Torres:2008yq}). It would allow to complete those diagrams with the close to 40 systems that now have occultation measurements but do not have parallaxes (Triaud et al. in prep).

\section{Conclusion}

The launch of ESA's {\it GAIA} mission in December of this year will permit to increase the number of system with parallaxes. With additional systems we may start seeing the emergence of {\it planet families}. For example, we could check whether GJ\,436 and HD\,209458b are real outliers, or the first representatives of sub-classes of objects. Just like the compilation of the Herzsprung-Russell diagram inspired theoretical developments in stellar evolution, theoretical sequences in colour-magnitude space could be computed for extrasolar planets. For known gravity, size and impacting flux, it may provide the means to distinguish different compositions and may even provide tests for the presence of an atmosphere, an ocean, or a surface \citep{Selsis:2011lr}, which has implications in our eventual selection of suitable targets for the search for life on other worlds.

\section*{Acknowledgments}

This work originates from an inspiring discussion with Jackie Faherty in a sandwich shop in Santiago de Chile (called La Superior, on Nueva de Lyon), during which for the first time I really looked at a colour-magnitude diagram. I would like to thank Michael Gillon for reading over,  to Heather Knutson for providing the phase curve data of HD\,189733b, and to Nikole Lewis and Pierre Maxted for theirs. I would like to thank the anonymous referee for her/his comments that led to an improvement of the paper.
I have made an extensive use of the \href{http://exoplanet.eu}{exoplanet.eu} \citep{Schneider:2011lr} and \href{http://exoplanets.org}{exoplanets.org} \citep{Wright:2011fj} websites, as well of the NASA/ADS, Simbad and VizieR repositories. 

This publication used data products from the Wide-field Infrared Survey Explorer, which is a joint project of the University of California, Los Angeles, and the Jet Propulsion Laboratory/California Institute of Technology, funded by the National Aeronautics and Space Administration. Data from the Two Micron All Sky Survey was also employed, which is a joint project of the University of Massachusetts and the Infrared Processing and Analysis Center/California Institute of Technology, funded by the National Aeronautics and Space Administration and the National Science Foundation.

I received funding under the from of a fellowship provided by the Swiss National Science Foundation under grant number PBGEP2-145594. During this fellowship, I am hosted by Joshua Winn, at the MIT Kavli institute and would like to thank him particularly, but also my colleagues for their welcome.
\bibliographystyle{mn2e}
\bibliography{../1Mybib.bib}

\bsp

\label{lastpage}

\end{document}